\documentclass[onecolumn,showpacs,preprintnumbers,amsmath,amssymb,pre,showkeys]{revtex4-1}

\usepackage{graphicx}
\usepackage{amsmath}
\usepackage{bm}
\usepackage{dcolumn}

\begin{document}

\newcommand{\p}{\partial}

\newcommand{\bea}{\begin{eqnarray}}
\newcommand{\eea}{\end{eqnarray}}
\newcommand{\be}{\begin{equation}}
\newcommand{\ee}{\end{equation}}
\newcommand{\bse}{\begin{subequations}}
\newcommand{\ese}{\end{subequations}}
\renewcommand{\l}{\ell}
\renewcommand{\p}{\partial}
\renewcommand{\ss}[1]{_{\hbox{\tiny #1}}}
\newcommand{\hk}{h_{\hbox{\tiny k}}}
\renewcommand{\bm}[1]{\mathbf{#1}}

\title{
Kink dynamics with oscillating forces
}

\author{Thomas Le Goff$^{1}$, Olivier Pierre-Louis$^1$, and  Paolo Politi$^{2,3}$}
\affiliation{
$^1$ Institut Lumi\`ere Mati\`ere, UMR5306 Universit\'e Lyon 1-CNRS, Universit\'e de Lyon 69622 Villeurbanne, France\\
$^2$ Istituto dei Sistemi Complessi, Consiglio Nazionale delle Ricerche, Via Madonna del Piano 10, 50019 Sesto Fiorentino, Italy\\
$^3$ INFN Sezione di Firenze, via G. Sansone 1, 50019 Sesto Fiorentino, Italy.
}

\date{\today}

\begin{abstract}
It is well known that the dynamics of a one-dimensional dissipative system driven by the Ginzburg-Landau free energy
may be described in terms of interacting kinks: two neighbouring kinks at distance $\l$ feel 
an attractive force $F(\l)\approx\exp(-\ell)$.
This result is typical of a bistable system whose inhomogeneities
have an energy cost due to surface tension, but for some physical systems
bending rigidity rather than surface tension plays a leading role. 
We show that a kink dynamics is still applicable, but the force $F(\l)$ 
is now oscillating, therefore producing configurations which are locally stable.
We also propose a new derivation of kink dynamics, which  
applies to a generalized Ginzburg-Landau free energy with an arbitrary combination
of surface tension, bending energy, and higher-order terms.
Our derivation is not based on a specific multikink approximation and the
resulting kink dynamics reproduces correctly the full dynamics of the original model.
This allows to use our derivation with confidence
in place of the continuum dynamics, reducing simulation time by orders of magnitude.
\end{abstract}



\maketitle

\section{Introduction}

The continuum description of a physical system requires to define a suitable,
coarse grained order parameter $h(\mathbf{x},t)$ and to build a free energy ${\cal F}$ if the system is at equilibrium,
or a partial differential equation (PDE)
obeyed by $h$ if the system is out of equilibrium.
In some cases, the PDE itself can be derived by some free energy. This is surely the case for
a system relaxing towards equilibrium (think to a phase separation process~\cite{Bray}), but it may
also be true for pattern forming systems, in which case ${\cal F}$ is a
{\it pseudo} free energy (also called Lyapunov functional~\cite{HH}).

Typically, ${\cal F}$ is made up of a potential part $\tilde U(h)$, which
 is the energy density for an homogeneous state, plus a part which accounts
for the energy cost of the inhomogeneities of the order parameter. 
The simplest way to weight spatial variations of $h(\bm{x},t)$ is to consider a term proportional to 
$(\nabla h)^2$. This surface tension term appears in completely different contexts,
from magnetism to surface physics. In the former case,
the misalignment of spins produces an energy cost which is
proportional to the gradient square of the magnetization~\cite{Langer}.
In the latter case, if the energy of a surface of local height $h(\bm{x})$ is
proportional to the total extension of the surface, we simply get $S=
\int d{\bm x}\sqrt{1+(\nabla h)^2}\simeq
S_0 +\frac{1}{2} \int d{\bm x} (\nabla h)^2$,
where $S_0=\int d{\bm x}$ is the area of the system.

If surface tension combines with a double well potential $\tilde U(h)$, which accounts
for the existence of two macroscopic stable states, ${\cal F}={\cal F}\ss{GL}$ is called 
Ginzburg-Landau free energy and it plays a relevant role in the theory of phase transitions and phase ordering.
In one dimension, a simple description of energetics and dynamics can be given in terms of kinks~\cite{Nepo}. 
A kink $\hk(x)$ is
the simplest non-homogeneous state which interpolates between the two minima of the potential,
$\pm h\ss{m}$,
and it has two main features: it is a monotonous function, and it is localized,
i.e. its derivative is exponentially small except in a finite size
region. The explicit expression of a kink for a specific potential, see Eq.~(\ref{eq_kink}), 
$\hk(x)=h\ss{m}\tanh(h\ss{m}x/\sqrt{2})$, make both properties obvious.

The reason why kinks play a major role derives from
the possibility to describe $h(x,t)$ as a sequence of kinks and,
finally, by the possibility to describe the continuum dynamics in terms
of an effective dynamics between kinks, which act as fictitious, interacting particles.
In poor terms, kinks have an attractive force which decreases exponentially with their distance:
the attractive force implies instability and coarsening;
the exponential dependence with distance implies coarsening is logarithmically slow.

In spite of the widespread importance of ${\cal F}\ss{GL}$,
we should not come to the wrong conclusion that its form is universal.
This caveat is particularly appropriate if bending rigidity is important:
soft matter and biophysics, dealing with membranes~\cite{membranes} and filaments~\cite{filaments},
are a relevant example.
This fact, the relevance of bending rigidity with respect to surface tension,
is not purely phenomenological. On the contrary, it has been recently
derived rigorosuly from an hydrodynamic model~\cite{Thomas_PRE}.
According to this model, the energy cost of inhomogeneities is proportional
(in a one-dimensional model) to the squared second spatial derivative of $h$, $(h_{xx})^2$, 
rather than to the squared derivative, $(h_x)^2$.
This modification is of paramount importance, because kinks are no longer
monotonous functions and this fact will be seen to change drastically their dynamics,
which turns out to be frozen.

The goal of our manuscript is twofold:
first, we extend ${\cal F}\ss{GL}$ to a free energy which depends on surface tension,
bending and possibly higher order terms.
Second, we reconsider the problem to pass from a continuos formulation
of the dynamics to a discrete description in terms of kinks,
proposing a new approach. A detailed numerical comparison
with continuum dynamics reveals that standard approaches where the
order parameter profile is written as a superposition of kinks fail to
reproduce quantitatively exact dynamics. Instead, our new approach 
is quantitative.

The paper is organized as follows. In Section II we define the various continuos models
and in Section III we give a simple derivation of known results. 
In Section IV we propose a new derivation of kink dynamics and compare numerically
different approaches.
In Section V we discuss the stability of steady states and in Section VI we
summarize the results.

\section{Continuous models}

As explained in the Introduction, a good starting point to introduce the dissipative dynamics of interest for us here is
the Ginzburg-Landau free energy.
For a scalar order parameter in a one-dimensional system, 
\be
\tilde{\cal F}\ss{GL}=\int dx \left( \frac{K_1}{2} h_x^2 + \tilde U(h)\right),
\label{F_gl}
\ee
where $\tilde U(h)$ is an arbitrary symmetric double well potential, 
with two equivalent minima for $h=\pm h\ss{m}$, which are the
ground states of the full free energy.
If $\tilde U(h)=U_0 U(h)$, rescaling space we obtain
\be
\tilde{\cal F}\ss{GL}  
=\sqrt{K_1 U_0} \int dx \left(\frac{1}{2} h_x^2 + U(h)\right) \equiv
e_0 {\cal F}\ss{GL}.
\ee
In the following the energy scale $e_0$ will be set equal to one, while
we don't rescale $h\ss{m}$ to one for pedagogical reasons.
Furthermore, for definiteness, in this Section we consider a standard
quartic potential, $U(h)=-h\ss{m}^2h^2/2+h^4/4$.

The free energy ${\cal F}\ss{GL}$ is the starting point to study the dissipative dynamics
when the system is relaxing towards equilibrium.
When studying dynamics the existence of conservation laws is of primary importance and
two main universality classes exist, depending on whether the order
parameter, $h(x,t)$, is conserved or not.
In the two cases we obtain, respectively, the Cahn-Hilliard (CH) and the
Time Dependent Ginzburg Landau (TDGL) equation,
\bea
\p_t h(x,t) &=& -\frac{\delta {\cal F}\ss{GL}}{\delta h}= h_{xx} + h^2\ss{m}h -h^3  \quad \mbox{TDGL}\label{TDGL},\\ 
\p_t h(x,t) &=& \p_{xx}\frac{\delta {\cal F}\ss{GL}}{\delta h} = -\p_{xx}(h_{xx} + h^2\ss{m}h -h^3)  \quad \mbox{CH}\quad \label{CH}.
\eea
In both cases, it is straightforward to show that
\be
\frac{d {\cal F}\ss{GL}}{dt} =\int dx \frac{\delta {\cal F}}{\delta h}\frac{\partial h}{\partial t} \le 0 .
\ee

Equation~(\ref{TDGL}) typically describes phase separation in a magnet,
because in this case relaxation dynamics does not conserve magnetization. 
Equation~(\ref{CH}) can instead describe phase separation in a binary alloy, where
matter 
is conserved. 
Here we will focus to a so-called {\it symmetric} quench, where the average value of 
the order parameter is zero.

In the above two cases, TDGL and CH equations, the overall picture of dynamics is well known.
The solution $h=0$, corresponding to the disordered 
or homogeneous phase, is linearly unstable, as easily seen by a stability analysis.
In fact, if $h(x,t)=\varepsilon e^{\sigma t}e^{iqx}$, to first order in $\varepsilon$ we find
\be
\sigma(q) = \left\{ 
\begin{array}{cc}
h\ss{m}^2-q^2 & \qquad \mbox{TDGL}\\
h\ss{m}^2 q^2-q^4 & \qquad \mbox{CH}
\end{array}
\label{sigma}
\right. ,
\ee
so that the homogeneous solution is linearly unstable for small $q$.
Because of such instability, small regions of the two phases $h=\pm h\ss{m}$ appear,
separated by kinks.
A kink is a steady solution of TDGL/CH equations which connects the two minima 
of the potential $U(h)$ for $x\to\pm\infty$. For the standard quartic potential,
such solution has the simple form
\be
h(x)=\pm h\ss{k}(x)\equiv \pm h\ss{m}\tanh\left(\frac{h\ss{m}}{\sqrt{2}} x\right). 
\label{eq_kink}
\ee

More generally, TDGL/CH equations have periodic solutions of arbitrarily large
wavelength which can be thought of as superpositions of
kinks $(h\ss{k}(x))$ and antikinks $(-h\ss{k}(x))$.
These kinks feel an attractive interaction, and 
when a kink and an antikink meet they annihilate,
therefore leading to an increasing average distance between the remaining ones (coarsening process).
In an infinite system this process lasts forever, but in one dimension it is logarithmically slow.

The above picture is well known and goes back to works by Langer~\cite{Langer} and Kawasaki and 
Ohta~\cite{Kawasaki_Ohta}. 
The main idea is to write $h(x,t)$ as a suitable superposition of positive and negative kinks,
getting a set of discrete equations for their positions $x_n(t)$.
This approach will be discussed in the next Section.
First, we need to show how this picture should be modified if the surface tensione term
($h_x^2$) in the GL free energy is replaced by a bending term ($h_{xx}^2$).

If bending rigidity dominates
over surface tension, the Ginzburg-Landau free energy should be written
\be
{\cal F}\ss{GL4}=\int dx \left[ \frac{1}{2} h_{xx}^2 + U(h) \right] ,
\label{F_gl4}
\ee
and Eqs.~(\ref{TDGL},\ref{CH}) are modified as follows,
\bea
\p_t h(x,t) = -h_{xxxx} + h\ss{m}^2 h - h^3 \quad\mbox{TDGL4}, \label{TDGL4}\\
\p_t h(x,t) = -\p_{xx}\left( -h_{xxxx} + h\ss{m}^2 h - h^3\right) \quad\mbox{CH4}, \label{CH4}
\eea
where the label `4' highlights the replacement of a second spatial derivative with a forth
spatial derivative. 
In its turn, the linear spectra (\ref{sigma}) should be replaced by
\be
\sigma(q) = \left\{ 
\begin{array}{cc}
h\ss{m}^2-q^4 & \qquad \mbox{TDGL4}\\
h\ss{m}^2 q^2-q^6 & \qquad \mbox{CH4}
\end{array}
\right. ,
\ee
showing that the homogeneous state is still unstable for large wavelength fluctuations.

\begin{figure}
\begin{center}
\includegraphics[width=6cm]{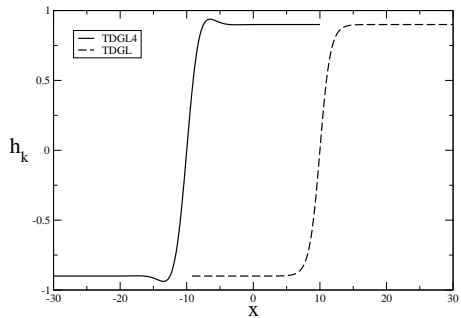}
\end{center}
\caption{
Plot of kinks appearing in TDGL/CH (dashed line) and in TDGL4/CH4 (full line).
In the latter case, the tail continues to oscillate around $\pm h\ss{m}$, but its exponential decay
allows to make visible only the first two oscillations.
}
\label{fig_kinks}
\end{figure}

In spite of these similarities, the study of steady states is not straightforward as for TDGL/CH, where it essentialy boils down
to solve the problem of a particle of coordinate $h$ in the potential $V(h)=-U(h)$.
Steady states are now determined by the time independent equation 
\be
-h_{xxxx} -U'(h)=0.
\label{eq_kink4}
\ee
The forth order derivative introduces new classes of kinks, because fixing the conditions
$h(x\to\pm\infty)=\pm h\ss{m}$ is no longer sufficient to uniquely determine a solution.
According to Ref.~\cite{Peletier_Troy} kinks can be labeled by their number of zeros, i.e. the number of
points where the kink profile vanishes (Eq.~(\ref{eq_kink}) shows that for TDGL/CH kinks
this number is equal to one).
The asymptotic behavior, i.e. the limiting behavior of $h\ss{k}(x)$ for large $|x|$, 
is determined by the linearization of Eq.~(\ref{eq_kink4}) around $h=h\ss{m}$,
\be
h_{xxxx} = -U''(h\ss{m}) (h-h\ss{m}).
\ee
It is easily found that $h(x)=h\ss{m}+R(x)$, where the tail $R(x)$ is given by
\be
R(x) = A\cos\left(\kappa x+\alpha\right)\exp\left(-\kappa x\right),
\label{R4}
\ee
where $\kappa=(U''(h\ss{m}))^{1/4}/\sqrt{2}$, while
the amplitude $A$ and the phase $\alpha$ are undetermined within the linear theory.
The exact shape of kinks for TDGL and TDGL4 models is plotted in Fig.~\ref{fig_kinks},
where we limit for TDGL4 to the kink with only one zero.

A similar picture, oscillating kinks and kinks with more zeros, emerges in other
PDEs, e.g. the convective Cahn-Hilliard equation~\cite{Zaks}.
In both cases there is no evidence of such multihump kinks during dynamics,
which lead us to assume they are dynamically irrelevant.
Therefore in the next Section we are studying kink dynamics assuming kinks which
cross the horizontal axis only once.

\section{Kink dynamics made simple}
\label{sec_skd}

The following, semiquantitative treatment of a profile simply consisting of the superposition of
a negative and a positive kink allows to grasp the relation between the kink tail $R(x)$ and
the kink interaction. 
In order to get a result as general as possible, we consider an energy functional
which is the sum of a symmetric double well potential (as before) plus
arbitrary quadratic terms,
whose only constraint is to satisfy the symmetry $x\to -x$.
Its most general form is
\be
{\cal F}=\int dx \left[U(h)-\frac{1}{2}\sum_{i=0}^M (-1)^i a_{2i}(\partial^i_x h)^2\right] ,
\label{F_ggl}
\ee
where $a_{2i}$ are constants and the notation $(\partial^i_x h)$ means the $i-$th order spatial derivative
of $h$.
We have also introduced the factor $\frac{1}{2}(-1)^i$ so as to get rid of it when evaluating
the functional derivative, according to the relations
\be
\begin{split}
\frac{\delta {\cal F}}{\delta h} 
= &
U'(h) - \sum_{j=1}^M a_{2j}\partial_{x}^{2j}h ,\\
\equiv & U'(h) - {\cal L}[h].
\end{split} \ee

The model we are going to analyze is a nonconserved, purely dissipative model,
where dynamics is driven by ${\cal F}$ according to the relation 
$\p_t h = -(\delta {\cal F}/\delta h)$, i.e.,
\be
\p_t h = {\cal L}[h] - U'(h).
\label{TDGL2-4}
\ee

If $\hk(x)$ is the kink profile centred at $x=0$, the two-kinks approximation amounts to writing
\be
h(x,t)=\hk(x+x_0(t)) - \hk(x-x_0(t)) -h\ss{m},
\label{2ka}
\ee
where the kinks are centred in $\pm x_0(t)$ and the constant term must be added
in order to get the correct values in the different regions (for an $N-$kinks approximation,
the constant term is more complicated, see Eq.~(2.8) of Ref.~\cite{Kawasaki_Ohta} and Eq.~(\ref{app_multikink})
here below).
Using Eq.~(\ref{2ka}) it is easy to evaluate $\p_t h$,
\be
\p_t h(x,t) = \dot x_0 \big(h'\ss{k}(x+x_0) + h'\ss{k}(x-x_0)\big),
\ee
 and its spatial integration,
\be
\int_{-\infty}^{+\infty} dx \p_t h(x,t) = 4h\ss{m}\dot x_0.
\label{x0p}
\ee
As for the RHS of Eq.~(\ref{TDGL2-4}), 
while we simply have 
\be
{\cal L}[h]={\cal L}[\hk(x+x_0)]-{\cal L}[\hk(x-x_0)],
\label{L_hk}
\ee
the evaluation of $U'(h)$ is a bit more involved. As soon as $|x|\gg a$, $a$ being
the size of the core of the kink,
$\hk(x)\simeq \pm [ h\ss{m} + R(|x|)]$, for $x\gtrless 0$ respectively.
Therefore, we can approximate Eq.~(\ref{2ka}) as follows
\be
h(x,t) \simeq \left\{
\begin{array}{cr}
\hk(x+x_0) +R(-x+x_0) & \quad\mbox{for } x<0,\\
-\hk(x-x_0) +R(x+x_0) & \quad\mbox{for } x>0,
\end{array}
\right.
\ee
and write, in the two cases,
\be
U'(h) \simeq \left\{
\begin{array}{c}
U'(\hk(x+x_0)) +U''(\hk(x+x_0))R(-x+x_0) \qquad\mbox{for } x<0 \\
-U'(\hk(x-x_0)) +U''(\hk(x-x_0))R(x+x_0) \qquad\mbox{for } x>0
\end{array}
\right. ,
\ee
so that
\begin{equation} \begin{split}
& \int_{-\infty}^{+\infty} dx U'(h) =
\int_{-\infty}^{+\infty} dx\left[ U'(\hk(x+x_0)) - U'(\hk(x-x_0)) \right] \\
&+ \int_{-\infty}^0 dx \left[ U'(\hk(x-x_0)) + U''(\hk(x+x_0))R(-x+x_0) \right] \\
&+ \int_0^{+\infty} dx \left[ -U'(\hk(x+x_0)) + U''(\hk(x-x_0))R(x+x_0) \right] . 
\end{split} 
\label{int_U}
\end{equation}
In the previous expression, a simple change of variable in the
second line integral, $x\to -x$, shows it is equal to the third line integral.

We can now match the spatial integration of the two sides of Eq.~(\ref{TDGL2-4}).
Using Eqs.~(\ref{x0p},\ref{L_hk},\ref{int_U}), we obtain
\be\begin{split}
4h\ss{m}\dot x_0 =& \int_{-\infty}^{+\infty} dx \big( {\cal L}[h] - U'(h)\big) \\
=& \int_{-\infty}^{+\infty} dx \Big( {\cal L}[\hk(x+ x_0)] -U'(\hk(x+ x_0)) \Big)\\
-& \int_{-\infty}^{+\infty} dx \Big( {\cal L}[\hk(x- x_0)] -U'(\hk(x- x_0)) \Big) \\
+& 2\int_{0}^{+\infty} dx \Big( U'(\hk(x+x_0)) - U''(\hk(x-x_0))R(x+x_0) \Big). 
\end{split} \ee
Since the integrands in the second and third line vanish, we finally get
\be \begin{split}
\dot x_0 =& 
\frac{1}{2h\ss{m}}\int_0^\infty dx\Big( U'(\hk(x+x_0)) -U''(\hk(x-x_0))R(x+x_0)\Big)\\
\simeq & \frac{1}{2h\ss{m}}\int_0^\infty dx\Big( U'(h\ss{m}+R(x+x_0)) -U''(\hk(x-x_0))R(x+x_0)\Big)\\
=& \frac{1}{2h\ss{m}}\int_0^\infty dx \Big(U''(h\ss{m})-U''(\hk(x-x_0))\Big) R(x+x_0) + {\cal O}(R^2).
\end{split} \ee
The quantity within large brackets in the final integral is exponentially small when $|x-x_0|\gg a$, 
so we can approximate the integral as the integrand value for $x=x_0$ times
the extension over which the function in square brackets is non vanishing, i.e. $a$.
Finally, we can write
\be
\dot x_0 \simeq \frac{a}{2h\ss{m}}[U''(h\ss{m})-U''(0)] R(\ell),
\label{s2kd}
\ee
with $\ell=2x_0$.
In conclusion, the speed of the right kink is barely proportional to $R(\ell)$, where $\ell$ is its
distance from the left kink (the quantity in square brackets being positive,
since $U''(h\ss{m})>0$ and $U''(0)<0$).

This result means that
a kink exerts a force on its right neighbour at distance $\ell$, force which is proportional to $R(\ell)$, where $R(x)$ is
the difference between the kink profile and its limiting value for large, positive $x$,
$R(x) = \hk(x) -h\ss{m}$. For the standard TDGL
equation, $\hk(x)=h\ss{m}\tanh(\frac{h\ss{m}}{\sqrt{2}}x)$ and $R(x)=R_2(x)$, with
\be
R_2(x) = -2h\ss{m}\exp(-\frac{h\ss{m}}{\sqrt{2}}x),
\label{R2}
\ee
while for TDGL4, $R(x)=R_4(x)\equiv A\cos\left(\kappa x+\alpha\right)\exp\left(-\kappa x\right)$,
see Eq.~(\ref{R4}). 

We can assume that Eq.~(\ref{s2kd}) may generalize to any sequence of
kinks located in $x_n(t)$ (with $x_{n+1}>x_n$),
\be
\dot x_n = \frac{1}{\hk'(0)}\left( U''(h\ss{m})-U''(0)\right) [R(x_n-x_{n-1}) - R(x_{n+1}-x_n)],
\label{snkd}
\ee
where the size $a$ of the kink core has been evaluated as $a=2h\ss{m}/\hk'(0)$.
Above equation should be supplemented by
the constraint that two neighbouring kinks annihilate when they overlap
(see details on numerical schemes in Appendix~\ref{app_num}).

As a matter of fact, such kink dynamics can be derived using 
a superposition of $N$ kinks,
\be
\begin{split}
h(x,t) &= (-1)^n \hk(x-x_n(t)) \\
&+ \sum_{k<n} (-1)^k [\hk(x-x_k(t)) - h\ss{m}]\\ 
&+\sum_{k>n} (-1)^k [\hk(x-x_k(t)) + h\ss{m}].
\label{app_multikink}
\end{split}
\ee
This approach was initially used by Kawasaki and Ohta~\cite{Kawasaki_Ohta} to study TDGL and CH equations.
In the next Section we are going to propose a novel approach and to compare both with numerical
integration of the full continuum equations.

\section{Improved kink dynamics}
\label{sec_kd}

We now provide a more general approach to kink dynamics:
we don't assume explicitely a specific ``multikink" approximation, as, e.g., Eq.~(\ref{app_multikink}),
and we consider the general energy functional given in Eq.~(\ref{F_ggl}).
We don't claim our approach is rigorously founded: its validity (and usefulness) 
are rather supported by the final comparison with numerics.

\subsection{Nonconserved case}

The nonconserved case corresponds to the dynamics
\begin{equation}
\partial_{t}h= -\frac{\delta {\cal F}}{\delta h} = \sum_{i}a_{2i}\partial_{x}^{2i}h-U^{\prime}(h).
\label{nc}
\end{equation}

\begin{figure}
\begin{center}
\includegraphics[height=2cm]{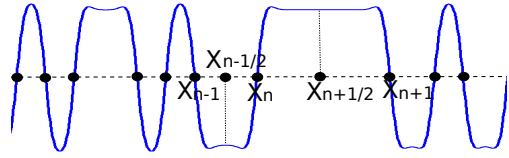}
\end{center}
\caption{
(Color online)
Schematic of studied system with relevant notations.
}
\label{schematic}
\end{figure}

In Fig.~\ref{schematic} we show the schematic of the system. It has been drawn for TDGL4/CH4 kinks, but notations are
generally valid. More precisely, $x_n$ means the position of $n-$th kink and
$x_{n\pm\frac{1}{2}}$ the points halfway between kinks $n$ and $(n\pm 1)$.
For ease of notation, $x_{n\pm\frac{1}{2}}$ is replaced by $n\pm\frac{1}{2}$ in integrals' extrema
and $h(x_{n\pm\frac{1}{2}})$ is replaced by $h_{n\pm\frac{1}{2}}$.

\begin{figure}
\begin{center}
\includegraphics[height=5cm]{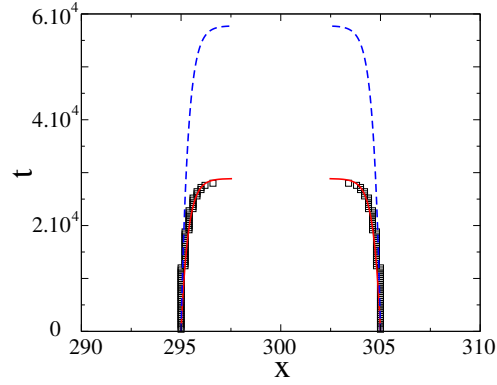}
\end{center}
\caption{
(Color online)
Exact dynamics and analytical approximations of the motion of two kinks for the TDGL model.
Black squares: exact dynamics (integration of Eq.~(\ref{TDGL})). Red full line: our model, Eq.~(\ref{v_n-tdgl}),
and Ei and Ohta's model. Blue dashed line: Kawasaki and Ohta's model.
}
\label{fig_tdgl}
\end{figure}

\begin{figure}
\begin{center}
\includegraphics[height=5cm]{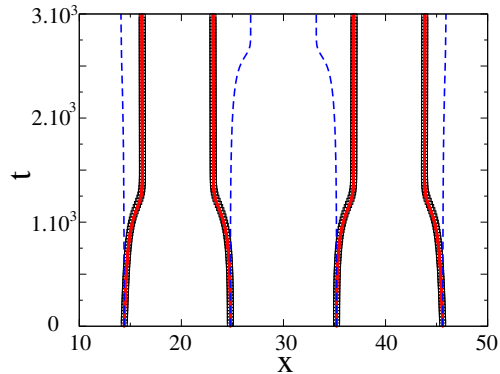}
\end{center}
\caption{
(Color online)
Exact dynamics and analytical approximations of the motion of four kinks for the TDGL4 model.
Black squares: exact dynamics (integration of Eq.~(\ref{TDGL4})). Red full line: our model, Eq.~(\ref{v_n-tdgl4}).
Blue dashed line: Kawasaki and Ohta's model.
}
\label{fig_tdgl4}
\end{figure}

\begin{figure}
\begin{center}
\includegraphics[height=5cm]{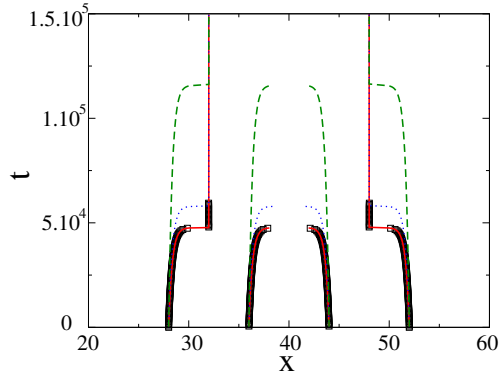}
\end{center}
\caption{
(Color online)
}
Exact dynamics and analytical approximations of the motion of four kinks for the CH model.
Black squares: exact dynamics (integration of Eq.~(\ref{CH})). Red full line: our model, Eq.~(\ref{v_n-ch}).
Blue dotted line: our model, Eq.~(\ref{v_n-ch_simple}). 
Green dashed line: Kawasaki and Ohta's model.
\label{fig_ch}
\end{figure}

\begin{figure}
\begin{center}
\includegraphics[height=5cm]{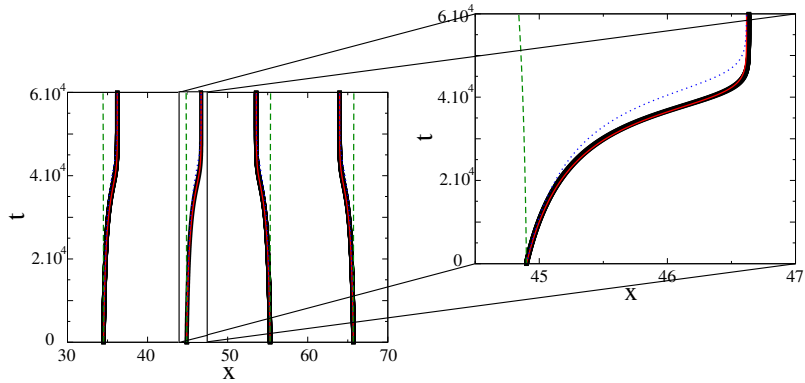}
\end{center}
\caption{
(Color online)
Exact dynamics and analytical approximations of the motion of four kinks for the CH4 model.
Black squares: exact dynamics (integration of Eq.~(\ref{CH4})). Red full line: our model, Eq.~(\ref{v_n-ch4}).
Blue dotted line: our model, Eq.~(\ref{v_n-ch4_simple}).
Green dashed line: Kawasaki and Ohta's model.
}
\label{fig_ch4}
\end{figure}

We assume that apart from the annihilation process, which occurrs when 
$\ell_n = x_{n+1}-x_n \approx a$, 
kinks retain their profile when moving. So, for $x$ around $x_n$
the previous equation can be rewritten as
\begin{equation}
-\dot{x}_{n}\partial_{x}h=\sum_{i}a_{2i}\partial_{x}^{2i}h-U^{\prime}(h).
\end{equation}
We then multiply both terms by $\partial_{x}h$ and integrate between $x_{n-\frac{1}{2}}$ and $x_{n+\frac{1}{2}}$ :
\begin{equation}
-\dot{x}_{n}\int_{n-\frac{1}{2}}^{n+\frac{1}{2}}\mathrm{d}x~(\partial_{x}h)^2=\sum_{i}a_{2i}\int_{n-\frac{1}{2}}^{n+\frac{1}{2}}\mathrm{d}x~\partial_{x}h\partial_{x}^{2i}h-U(h_{n+\frac{1}{2}})+U(h_{n-\frac{1}{2}}).\nonumber
\end{equation}
Direct integration and integration by parts give
\begin{equation}\label{kdnc}
\dot{x}_{n}=\frac{1}{\int_{n-\frac{1}{2}}^{n+\frac{1}{2}}\mathrm{d}x~(\partial_{x}h)^2}
\left[\sum_{i}a_{2i}\left(\frac{(-1)^{i}}{2}[(\partial_{x}^{i}h)^2]_{n-\frac{1}{2}}^{n+\frac{1}{2}}
+\sum_{k=1}^{k<\frac{i}{2}}[\partial_{x}^{2k}h\partial_{x}^{2i-2k}h]_{n-\frac{1}{2}}^{n+\frac{1}{2}}\right)
+U(h_{n+\frac{1}{2}})-U(h_{n-\frac{1}{2}})
\right].
\end{equation}
We stress that above result derives from one single assumption, $\p_t h\simeq -\dot{x}_{n}\partial_{x}h$ for 
$x$ close to $x_n$. 
Equation (\ref{kdnc}) can be further elaborated because in the region halfway between $x_n$ and $x_{n+1}$
we can expand $h(x,t)$ around the asymptotic values $\pm h\ss{m}$,
\be
h(x,t) \simeq \pm \left[ h\ss{m} + R(x-x_n)+R(x_{n+1}-x)\right] ,
\label{h_expansion}
\ee
where $+/-$ applies for a positive/negative $n-$th kink.
Using this notation, we finally get
\begin{equation}
\begin{split}
\dot{x}_{n}=&\frac{1}{\int_{-\infty}^{+\infty}\mathrm{d}x~(\partial_{x}h\ss{k})^2}
\left\{\sum_{i}a_{2i}\left[(1+(-1)^{i})\left(R^{(i)}\left(\frac{\l_{n}}{2}\right)^{2}-R^{(i)}\left(\frac{\l_{n-1}}{2}\right)^{2}\right)\right.\right.\\
&\left.\left.-4\sum_{k=1}^{k<\frac{i}{2}}\left(R^{(2k)}\left(\frac{\l_{n}}{2}\right)R^{(2i-2k)}\left(\frac{\l_{n}}{2}\right)-R^{(2k)}\left(\frac{\l_{n-1}}{2}\right)R^{(2i-2k)}\left(\frac{\l_{n-1}}{2}\right)\right)\right]+2U''(h\ss{m})\left(R^{2}\left(\frac{\l_{n}}{2}\right)-R^{2}\left(\frac{\l_{n-1}}{2}\right)\right)\right\} ,
\end{split}
\label{kdnc2}
\end{equation}
where, at denominator of Eq.~(\ref{kdnc}), we made the approximation
\be
\int_{n-\frac{1}{2}}^{n+\frac{1}{2}}\mathrm{d}x~(\partial_{x}h)^2 \simeq
\int_{n-\frac{1}{2}}^{n+\frac{1}{2}}\mathrm{d}x~(\partial_{x}\hk)^2 \simeq
\int_{-\infty}^{+\infty}\mathrm{d}x~(\partial_{x}\hk)^2 ,
\ee
i.e., 
we have assumed that close to $x_n$ the kink profile is similar to the static profile $\hk(x)$ and
we have extended the extrema of the integral to $\pm\infty$,
because $\partial_{x}\hk$ is concentrated around $x_n$.

Therefore, in the general case of an equation with several terms $a_{2i}\ne 0$ the expression of the
speed of a kink is fairly complicated. One remark is in order:
the different contributions to the RHS of Eq.~(\ref{kdnc2}) are not proportional to $R(\l_n)$ and $R(\l_{n-1})$, as appearing
in the simple approach given in the previous Section. 
This point is better clarified by focusing on two explicit cases.

$\bullet$ For TDGL, the only nonvanishing term in the summation (\ref{nc}) is $a_2=1$, so Eq.~(\ref{kdnc2})
strongly simplifies to
\be
\dot x_n = \frac{2U''(h\ss{m})}{\int_{-\infty}^{+\infty}\mathrm{d}x~(\partial_{x}\hk )^2}
\left[ R^2\left(\frac{\l_n}{2}\right) - R^2\left(\frac{\l_{n-1}}{2}\right) \right] ,
\quad \mbox{[NEW approach]}
\label{newtdgl2}
\ee
which must be compared with Eq.~(\ref{snkd}), rewritten here for convenience: 
\be
\dot x_n = \frac{1}{\hk'(0)}\left( U''(h\ss{m})-U''(0)\right) [R(\ell_{n-1}) - R(\ell_n)],
\quad \mbox{[KO approach]}
\tag{\ref{snkd}}
\ee
where KO stands for Kawasaki and Otha~\cite{Kawasaki_Ohta}.

In the specific TDGL case $R(\l)$ is a simple exponential, 
so that
\be
R(\l)= \mbox{constant}\,\times\, R^2\left( \frac{\l}{2}\right) .
\ee
In conclusion, the new approach (\ref{newtdgl2}) and the old approach (\ref{snkd})
differ for the prefactor only. Let us work out the two prefactors for
the explicit expression $U(h)=-\frac{h^2}{2}+\frac{h^4}{4}$. 
Using Eq.~(\ref{eq_kink}) for the kink profile and Eq.~(\ref{R2}) for its tail (both with $h\ss{m}=1$), 
we find
\bea
\label{v_n-tdgl}
\dot{x}_{n}=& 12\sqrt{2}\left[\exp\left(-\sqrt{2}\l_{n}\right)-\exp\left(-\sqrt{2}\l_{n-1}\right)\right],
\quad \mbox{[NEW approach]} \\
\dot{x}_{n}=& \phantom{1}6\sqrt{2}\left[\exp\left(-\sqrt{2}\l_{n}\right)-\exp\left(-\sqrt{2}\l_{n-1}\right)\right].
\quad \mbox{[KO approach]} 
\label{v_n-old}
\eea

Equation~(\ref{v_n-tdgl}) agrees with Ei and Ohta~\cite{Ei_Ohta} and with Carr and Pego~\cite{Carr_Pego}.
These authors use a perturbative approach where the small parameter is
the extension of the domain wall defining the kink, but while Carr and Pego rely on the existence
of a Lyapunov functional, Ei and Ohta do not.
Instead, Eq.~(\ref{v_n-old}) agrees with Kawasaki and Ohta~\cite{Kawasaki_Ohta},
whose approach has been exemplified in Sec.~\ref{sec_skd}. 
In Fig.~\ref{fig_tdgl} we compare old (dashed line) and new (full line) approach 
with exact kink dynamics (squares), showing that the new approach is quantitatively correct.

$\bullet$
For TDGL4 equation, the two approaches give substantially different results, as we are going to show.
In Eq.~(\ref{kdnc}) we now have only the term $i=2$, with $a_4=-1$, and
\be
\dot x_n = \frac{2}{\int_{-\infty}^{+\infty}\mathrm{d}x~(\partial_{x}\hk )^2}
\left\{ -\left[ \left( R''\left( \frac{\l_n}{2}\right) \right)^2 - 
\left( R''\left( \frac{\l_{n-1}}{2}\right) \right)^2 \right]
+ U''(h\ss{m}) \left[ R^2\left(\frac{\l_n}{2}\right) - R^2\left(\frac{\l_{n-1}}{2}\right) \right] \right\}
\label{kdtdgl4}
\ee

Now, see Eq.~(\ref{R4}), $R(\l)=A\cos(\kappa\l+\alpha)\exp(-\kappa\l)$, so that (even up to a constant)
\be
R(\l)\ne R^2\left( \frac{\l}{2}\right) \quad \mbox{and} \quad R(\l)\ne R''^2\left( \frac{\l}{2}\right).
\ee

If we use the correct expression for $R(\l)$ we obtain
\begin{equation}
\dot{x}_{n}=\frac{2U^{\prime\prime}_{m}A^{2}}{\int_{-\infty}^{+\infty}\mathrm{d}x~(\partial_{x}\hk )^2}
\left[\cos\left(\kappa \l_{n}+2\alpha\right)\exp\left(-\omega \l_{n}\right)-\cos\left(\kappa \l_{n-1}
+2\alpha\right)\exp\left(-\kappa \l_{n-1}\right)\right] .
\label{v_n-tdgl4}
\end{equation}
In Fig.~\ref{fig_tdgl4}, we compare the full numerical solution of the continuum TDGL4 model (squares) with our results (Eq.~(\ref{v_n-tdgl4}), full line) 
and with results obtained with the multikink approximation (dashed line).
Our new approach of kink dynamics reproduces quantitatively very well the full numerical solution.
In addition, the results from the multikink ansatz approach 
cannot be corrected using a simple rescaling of time, as in the case of TDGL.

\subsection{Conserved case}

Similarly to the nonconserved case, we are going to consider the general model
\begin{equation}
\partial_{t}h=-\partial_{xx}\left(\sum_{i}a_{2i}\partial_{x}^{2i}h-U^{\prime}(h)\right) ,
\label{c}
\end{equation}
which requires more involved mathematics, whose details are partly given in Appendix~\ref{app_cal}. 
Here we provide the final result,
\begin{equation}
\dot{x}_{n}=\frac{1}{4 h\ss{m}^{2}\l_{n}\l_{n-1}-A_{n}(\l_{n}+\l_{n-1})}
\left\{\l_{n-1}\left[\dot{x}_{n+1}A_{n+1}+f(\l_{n+1}, \l_{n-1})\right]
+\l_{n}\left[\dot{x}_{n-1}A_{n-1}+f(\l_{n}, \l_{n-2})\right]\right\}
\label{eq_app}
\end{equation}
where 
\be
A_{n}=\int_{n-\frac{1}{2}}^{n+\frac{1}{2}}\mathrm{d}x~\partial_{x}h\int_{n-\frac{1}{2}}^{x}\mathrm{d}x^{\prime}~(h-h_{n-\frac{1}{2}})
\label{eq_An}
\ee 
and
\be
\begin{split}
f(x,y)=\sum_{i}a_{2i} & 
\left[  (1+(-1)^{i})\left(R^{(i)}\left(\frac{x}{2}\right)^{2}-R^{(i)}\left(\frac{y}{2}\right)^{2}\right) \right. \\
& \left. -4\sum_{k=1}^{k<\frac{i}{2}}\left(R^{(2k)}\left(\frac{x}{2}\right)R^{(2i-2k)}\left(\frac{x}{2}\right)
-R^{(2k)}\left(\frac{y}{2}\right)R^{(2i-2k)}\left(\frac{y}{2}\right)\right)\right]+2U^{\prime\prime}_{m}\left(R^{2}\left(\frac{x}{2}\right)-R^{2}\left(\frac{y}{2}\right)\right),
\end{split}
\label{eq_fxy}
\ee
which reduces to
\begin{equation}
\begin{split}
\dot{x}_{n}=& \frac{1}{4 h\ss{m}^{2}\l_{n}\l_{n-1}-2\sqrt{2}(\l_{n}+\l_{n-1})}
\left\{\l_{n-1}\left[2\sqrt{2}\dot{x}_{n+1}+8U^{\prime\prime}_{m}\left(\exp(-\sqrt{2}\l_{n+1})-
\exp(-\sqrt{2}\l_{n-1})\right)\right]\right.\\
&\left. +\l_{n}\left[2\sqrt{2}\dot{x}_{n-1}+8U^{\prime\prime}_{m}\left(\exp(-\sqrt{2}\l_{n})-
\exp(-\sqrt{2}l_{n-2})\right)\right]\right\} \qquad \mbox{[CH]}
\end{split}
\label{v_n-ch}
\end{equation}
for the CH equation, and to
\begin{equation}
\begin{split}
\dot{x}_{n}=&
\frac{1}{4 h\ss{m}^{2}\l_{n}\l_{n-1}-A(\l_{n}+\l_{n-1})} \times \qquad\qquad\qquad\qquad \mbox{[CH4]}\\
&\left\{\l_{n-1}\left[\dot{x}_{n+1}A+2U^{\prime\prime}_{m}A^{2}
\left(\cos\left(\kappa \l_{n+1}+2\alpha\right)\exp\left(-\kappa \l_{n+1}\right)
-\cos\left(\kappa \l_{n-1}+2\alpha\right)\exp\left(-\kappa \l_{n-1}\right)\right)\right]\right.\\
&\left.+\l_{n}\left[\dot{x}_{n-1}A+2U^{\prime\prime}_{m}A^{2}
\left(\cos\left(\kappa \l_{n}+2\alpha\right)\exp\left(-\kappa \l_{n}\right)
-\cos\left(\kappa \l_{n-2}+2\alpha\right)\exp\left(-\kappa \l_{n-2}\right)\right)\right]\right\},
\end{split}
\label{v_n-ch4}
\end{equation}
for the CH4 equation, with $A=\int_{-\infty}^{+\infty} dx (h\ss{m}^2-\hk^2)$.

The previous two equations are rather involved and the expressions for kink speeds $\dot x_n$
are coupled, see the terms proportional to $\dot x_{n\pm 1}$ on the Right Hand Side.
Since the terms proportional to $\dot x_n$ in the  Right Hand Side of Eq.~(\ref{v_n-ch4})
are smaller than the term $\dot x_n$ on the Left Hand Side by a factor $\sim 1/\ell_n$ for large $\ell_n$,
we may neglect them when $\ell_n\gg a$. 
Analogously, at denominators we can neglect the terms linear in $\ell$ with respect the terms
quadratic in $\ell$. Finally, we obtain
a simplified version of Eqs.~(\ref{v_n-ch},\ref{v_n-ch4}):
\begin{equation}
\begin{split}
\dot{x}_{n}=& \frac{1}{4 h\ss{m}^{2}\l_{n}\l_{n-1}}
\left[8\l_{n-1}U^{\prime\prime}_{m}\left(\exp(-\sqrt{2}\l_{n+1})-
\exp(-\sqrt{2}\l_{n-1})\right)\right.\\
&\left. +8\l_{n}U^{\prime\prime}_{m}\left(\exp(-\sqrt{2}\l_{n})-
\exp(-\sqrt{2}\l_{n-2})\right)\right] \qquad \mbox{[CH simplified]}
\end{split}
\label{v_n-ch_simple}
\end{equation}
and
\begin{equation}
\begin{split}
\dot{x}_{n}=&
\frac{1}{4 h\ss{m}^{2}\l_{n}\l_{n-1}} \times \qquad\qquad\qquad\qquad \mbox{[CH4 simplified]}\\
&\left\{\l_{n-1}\left[2U^{\prime\prime}_{m}A^{2}
\left(\cos\left(\kappa \l_{n+1}+2\alpha\right)\exp\left(-\kappa \l_{n+1}\right)
-\cos\left(\kappa \l_{n-1}+2\alpha\right)\exp\left(-\kappa \l_{n-1}\right)\right)\right]\right.\\
&\left.+\l_{n}\left[2U^{\prime\prime}_{m}A^{2}
\left(\cos\left(\kappa \l_{n}+2\alpha\right)\exp\left(-\kappa \l_{n}\right)
-\cos\left(\kappa \l_{n-2}+2\alpha\right)\exp\left(-\kappa \l_{n-2}\right)\right)\right]\right\}.
\end{split}
\label{v_n-ch4_simple}
\end{equation}
In Figure~\ref{fig_ch}, we compare the different approaches and the numerical solution of the continuum CH equation,
while in Fig.~\ref{fig_ch4} we do the same for the CH4 equation.
In both cases, exact numerical results are given by squares, our full analytical expressions
Eqs.~(\ref{v_n-ch},\ref{v_n-ch4}) are given by solid lines,
our simplified expressions Eqs.~(\ref{v_n-ch_simple},\ref{v_n-ch4_simple}) are given by dotted lines,
and the analytical expressions using multikink approximations are given by dashed lines.
The two figures clearly show that our full expressions (\ref{v_n-ch},\ref{v_n-ch4})
reproduce correctly numerics of the continuum model in both cases. 
The simplified model provides a reasonable result, but it is quantitatively inaccurate,
proving that the subdominant terms $\sim 1/\ell_n$ are relevant for the interkink distances $\ell_n$
used in the simulations of Fig.~\ref{fig_ch} and Fig.~\ref{fig_ch4}.
However, these subdominant terms should become negligible for larger
interkink distances $\ell_n$.

\section{Stability of steady states}

In Sec.~\ref{sec_skd} we have shown that
TDGL-kinks feel an attractive interaction
while TDGL4-kinks feel an oscillating interaction, even if in both cases $R(x)$ vanishes exponentially at large $x$.
This fact implies two important differences:
(i)~all TDGL steady configurations are uniform, $x_{n+1}-x_n=\ell$, while TDGL4 ones may be even disordered;
(ii)~all TDGL steady states are linearly unstable, while TDGL4 steady states may be stable or unstable.
Let us prove these statements.

We can rewrite Eq.~(\ref{snkd}) incorporating the positive prefactor at RHS in $t$,
\be
\dot x_n = R(x_n-x_{n-1}) - R(x_{n+1}-x_n),
\label{skdnc}
\ee
whose time independent solution is $R(\ell_n)=R(\ell_{n-1})~\forall n$, i.e. $R(\ell_n)=r$, with $\ell_n=x_{n+1}-x_n$.
For the standard TDGL model $R(x)$ is a monotonous function, so the equation $R(\l_n)=r$ has at most one solution. In practice, every
uniform configuration $\ell_n=\ell$ is stationary. Instead, for the TDGL4 model, the equation 
\be
R(\ell_n) \equiv A \cos(\kappa\l_n +\alpha) \exp(-\kappa\l_n) = r
\ee
has a number of solutions which increases when decreasing $|r|$, up to an infinite number of solutions for $r=0$.

As for the stability of a steady state, let us first focus on uniform configurations, i.e. all $\l_n=\l$. 
In order to study the linear stability of this configuration we need to perturb it, 
\be
\ell_n(t)=\ell +\epsilon_n(t), 
\ee
and determine the temporal evolution of the perturbations $\epsilon_n(t)\ll \ell$. Using Eq.~(\ref{skdnc}) we get 
\bea
\label{nc_gen_stability}
\dot \epsilon_n &=& 2R(\ell_n) -R(\ell_{n+1})-R(\ell_{n-1}) \\
&=& R'(\ell)\big( 2\epsilon_n-\epsilon_{n+1}-\epsilon_{n-1}\big), 
\eea
whose single harmonic solution is $\epsilon_n(t)=\exp(\omega t +iqn)$, with
\be
\omega(q) = 4R'(\ell) \sin^2\left(\frac{q}{2}\right).
\ee
We have stability (instability) if $R'(\ell) <0 (>0)$.
Since $R_2(x)$ is an increasing function, see Eq.~(\ref{R2}), 
any uniform configuration is unstable for TDGL.
This result leads to a perpetual coarsening dynamics~\cite{Langer}.
Instead, since $R_4(x)$ is oscillating also its derivative is oscillating and with 
varying $\ell$ we obtain stable steady states if $R'(\l)<0$
and unstable steady states if $R'(\l)>0$.

In the general case of a nonuniform steady state,
\be
\l_n(t) = \l^*_n+\epsilon_n(t) \qquad \mbox{with} \quad R(\l^*_n)=r,
\ee
Eq.~(\ref{nc_gen_stability}), which is still valid, gives
\be
\dot \epsilon_n = 2R'(\l^*_n)\epsilon_n -R'(\l^*_{n-1})\epsilon_{n-1} -R'(\l^*_{n+1})\epsilon_{n+1}.
\ee
The linear character of the equations allows to write $\epsilon_n(t) =e^{\sigma t} A_n$, getting 
\be
2R'(\l^*_n)A_n -R'(\l^*_{n-1})A_{n-1} -R'(\l^*_{n+1})A_{n+1} = \sigma A_n
\label{e:An}
\ee
but  the $n$-dependence of  $R'(\ell_n^*)$  prevents
the diagonalization with Fourier modes
($A_n \ne e^{iqn}$).


Multiplying Eq.(\ref{e:An}) with $R'(\ell_n^*)A_n^\dagger$, summing aver all $n$, and after some simple recombinations of the l.h.s., we obtain
\begin{eqnarray}
\sum_{n=1}^N |R'(\ell_{n+1}^*)A_{n+1}-R'(\ell_n^*)A_n|^2= \sigma \sum_{n=1}^N R'(\ell_n^*)|A_n|^2 ,
\label{e:quad_form}
\end{eqnarray}
which shows that eigenvalues $\sigma$ are real. Furthermore,
if all quantities $R'(\ell_n^*)$ have the same sign, $\sigma$ has the sign of $R'(\ell_n^*)$.
In particular,
any steady-state kink configuration with $R'(\ell_n^*)<0$ for all $n$ is stable. 
As a consequence,  $R'(\ell_n^*)<0$ for all $n$ is a sufficient condition
for stability, and there is an infinite number of stable configurations
in which the system can be trapped and stuck during the dynamics. 

If the quantities $R'(\ell_n^*)$ exhibit both positive and negative signs, Eq.~(\ref{e:quad_form}) does not allow to draw conclusions.
However, in the simple cases of a period-2 configuration, $\l^*_n=\l^*_{n+2}$, 
or a period-3 configuration, $\l^*_n=\l^*_{n+3}$,
we can prove that $R'(\ell_n^*)<0$ is also a necessary condition for stability.
Let's show it explicitly for the period-2 configuration. If
\be
\l^*_{2n} = \l\ss{s2} \qquad
\l^*_{2n+1} = \l\ss{s1},
\ee
we obtain two coupled equations which are solved assuming 
\be
A_{2n} = c_2 e^{i2nq} \qquad
A_{2n+1} = c_1 e^{i(2n+1)q} . 
\ee
The resulting eigenvalue equation is
\be
\sigma^2 -2 \left( R'(\l\ss{s1}^*) + R'(\l\ss{s2}^*) \right)\sigma +
4 R'(\l\ss{s1}^*) R'(\l\ss{s2}^*) \sin^2 q =0.
\ee
We have stability if both eigenvalues are negative, i.e.
\be
\mbox{stability} \quad \Leftrightarrow \quad
R'(\l\ss{s1}^*)<0 \;\; \mbox{and} \;\; R'(\l\ss{s2}^*)<0 .
\ee

\section{Summary and discussion}

Our paper studies kink dynamics deriving from a generalized Ginzburg-Landau free energy,
see Eq.~(\ref{F_ggl}).
The potential part of the free energy, $U(h)$, is the classical, symmetric double well potential,
typical of a bistable system. The ``kinetic" part of the free energy
is the sum of squares of order parameter derivatives of general order. 

The main motivation to study such free energy is that there are systems whose ``kinetic"
free energy is not given by surface tension, proportional to $(h_x^2)$, but rather to
bending energy, which is proportional to $(h_{xx}^2)$. Since the two terms are not mutually
exclusive, it is quite reasonable to consider the free energy
\be
{\cal F} = \int dx \left[ U(h) + \frac{K_1}{2} (\p_x h)^2 + \frac{K_2}{2} (\p^2_x h)^2 \right] .
\ee
Then, we have generalized previous expression to Eq.~(\ref{F_ggl}). However, even if our treatment
is valid in full generality, we have focused on two cases: $K_1=1,K_2=0$ and $K_1=0,K_2=1$, i.e. to pure
surface tension systems (to check existing results) and to pure bending systems (novel system
of specifical biophysical interest~\cite{Thomas_PRE}). 

Once ${\cal F}$ is given, we may derive a generalized Ginzburg-Landau equation, see Eq.~(\ref{nc}), 
or a generalized Cahn-Hilliard equation, see Eq.~(\ref{c}).
The standard approach to derive an effective kink dynamics is to assume a specific form of $h(x,t)$ as
a suitable superposition of kinks, $\hk(x-x_n(t))$, located in $x_n(t)$. This method has proved to be
fruitful, because it has allowed to explain coarsening dynamics of TDGL/CH 
models~\cite{Kawasaki_Ohta,Kawakatsu_Munakata,Nagai_Kawasaki_I,Nagai_Kawasaki_II,Nagai_Kawasaki_III},
to determine coarsening exponents, to study the effect
of a symmetry breaking term~\cite{PP_kinks}, and the effect of thermal noise.

However, the ability of the multikink approximation to reproduce quantitatively
the exact dynamics of the continuum model was already
questioned by Ei and Ohta~\cite{Ei_Ohta} for the TDGL model.
The failure of this goal is even more transparent when considering the bending energy,
i.e. the TDGL4/CH4 models.
In Figures~\ref{fig_tdgl}-\ref{fig_ch4} we make a detailed comparison of exact results (squares, derived from the direct
integration of the equation) with the standard multikink approximation (dashed lines)
and with our new results (full lines). The conclusion is that the new approach gives a reliable, discrete
description of the exact, continuous dynamics: see how full lines follow squares in all Figs.~\ref{fig_tdgl}-\ref{fig_ch4}.

We can still ask why we should derive an approximate kink dynamics if we have the full exact dynamics
of order parameter $h(x,t)$. There are several good reasons:
(i)~an analytical approach to nonlinear full dynamics is hard if not impossible;
(ii)~kink dynamics is easy to understand and analytical methods are feasible;
(iii)~numerical simulation of kink dynamics is far faster than the simulation of the full PDE.

In addition to be numerically reliable, some of our kink models (TDGL4/CH4) have the 
advantage of showing an oscillating tail $R(x) = \hk(x) - h\ss{m}$. This oscillation implies two
important features. Firstly, an oscillating tail means an oscillating force between kinks, as opposed to the
classical TDGL/CH models. Therefore, the long term dynamical scenario is not a coarsening scenario,
but the freezing in one of the many stable states~\cite{Thomas_PRE}. This can give rise to a consistency problem 
when we use the approximation $\l_n\gg a$ to derive kink dynamics. However, the approximation is expected to give
reasonable results even for not so far kinks and comparison with exact numerics supports such claim.
 
Secondly, an oscillating tail $R(x)$ is at the origin of a quantitative discrepancy between classical multikink
approaches and our approach. Using numerical simulations, we have shown that our approach provides much better quantitative results. 
For example, classical results for TDGL4 provide an interkink force 
proportional to $R(\l)$ while a force $F(\l)\approx R^2(\l/2)$ appears to be more appropriate.
If it were $R(\l)\simeq \exp(-\kappa\l)$, the two approaches would be equivalent, apart a rescaling of time.
Instead, if $R(\l)\simeq \cos(\kappa\l +\alpha)\exp(-\kappa\l)$ the two approaches are definitely different.

In this paper we have focused on the derivation of kink dynamics and on the quantitative comparison with
the exact dynamics of the PDE. The kink models for TDGL4 and CH4 are also considered in Ref.~\cite{Thomas_long}
where we specially use them for long time dynamics of the deterministic model and for any time
dynamics of the stochastic models. In fact, once we have proven (here) their quantitative reliability,
we can use them with confidence whenever the direct numerical integration of PDEs would be
too demanding in terms of CPU time. This is certainly the case if we require to go to very long times
or if we need to add stochastic noise to the equations.
Our evaluation of the simulation times for the PDE ($t\ss{PDE}$) and for the kink model ($t\ss{k}$)
allows to conclude that we gain four orders of magnitude, $t\ss{PDE}/t\ss{k}\approx 10^4$.

\acknowledgments

We wish to thank Xavier Lamy for usueful
insights about the stability analysis of kink arrays.
We also acknowledge support from
Biolub Grant No. ANR-12-BS04-0008.

\appendix

\section{Derivation of Eq.~(\ref{eq_app})}
\label{app_cal}

Let us rewrite Eq.~(\ref{c}),
\begin{equation}
\partial_{t}h=-\partial_{xx}\left(\sum_{i}a_{2i}\partial_{x}^{2i}h-U^{\prime}(h)\right)
\label{C}
\end{equation}
and we still suppose we can write $\partial_{t}h\simeq-\dot{x}_{n}\partial_{x}h$.
If we integer \eqref{C} twice between $x_{n-\frac{1}{2}}$ and $x$ we obtain
\begin{equation}
-\dot{x}_{n}\int_{n-\frac{1}{2}}^{x}\mathrm{d}x^{\prime}~(h-h_{n-\frac{1}{2}})=-\sum_{i}a_{2i}\partial_{x}^{2i}h+U^{\prime}(h)+j_{n-\frac{1}{2}}(x-x_{n-\frac{1}{2}})+\mu_{n-\frac{1}{2}},\nonumber
\end{equation}
with $j=\partial_{x}\left(\sum_{i}a_{2i}\partial_{x}^{2i}h-U^{\prime}(h)\right)$ and
$\mu=\sum_{i}a_{2i}\partial_{x}^{2i}h-U^{\prime}(h)$. 
Then we multiply by $\partial_{x}h$ and we integer between $x_{n-\frac{1}{2}}$ and $x_{n+\frac{1}{2}}$ :
\begin{equation}\label{eqint1}
-\dot{x}_{n}A_{n}=\left[-\sum_{i}a_{2i}\left(\frac{(-1)^{i-1}}{2}(\partial_{x}^{i}h)^2-\sum_{k=1}^{k<\frac{i}{2}}\partial_{x}^{2k}h\partial_{x}^{2i-2k}h\right)+U(h)\right]_{n-\frac{1}{2}}^{n+\frac{1}{2}}+j_{n-\frac{1}{2}}B_{n}+\mu_{n-\frac{1}{2}}[h]_{n-\frac{1}{2}}^{n+\frac{1}{2}}.
\end{equation}
where $A_{n}=\int_{n-\frac{1}{2}}^{n+\frac{1}{2}}\mathrm{d}x~\partial_{x}h\int_{n-\frac{1}{2}}^{x}\mathrm{d}x^{\prime}~(h-h_{n-\frac{1}{2}})$ and $B_{n}=\int_{n-\frac{1}{2}}^{n+\frac{1}{2}}\mathrm{d}x~\partial_{x}h(x-x_{n-\frac{1}{2}})$.

If we do the same thing but between $x$ and $x_{n+\frac{1}{2}}$ we obtain :
\begin{equation}\label{eqint2}
-\dot{x}_{n}A_{n}=\left[-\sum_{i}a_{2i}\left(\frac{(-1)^{i-1}}{2}(\partial_{x}^{i}h)^2-\sum_{k=1}^{k<\frac{i}{2}}\partial_{x}^{2k}h\partial_{x}^{2i-2k}h\right)+U(h)\right]_{n-\frac{1}{2}}^{n+\frac{1}{2}}+j_{n+\frac{1}{2}}B^{\prime}_{n}+\mu_{n+\frac{1}{2}}[h]_{n-\frac{1}{2}}^{n+\frac{1}{2}},
\end{equation}
where $B^{\prime}_{n}=-\int_{n-\frac{1}{2}}^{n+\frac{1}{2}}\mathrm{d}x~\partial_{x}h(x_{n+\frac{1}{2}}-x)$. 
By summing Eq.~\eqref{eqint1} for the $n-$th kink and Eq.~\eqref{eqint2} for the $(n-1)-$th kink we find

\begin{equation}\label{eqint3}
\begin{split}
-\dot{x}_{n}A_{n}-\dot{x}_{n-1}A_{n-1}=&\left[-\sum_{i}a_{2i}\left(\frac{(-1)^{i-1}}{2}(\partial_{x}^{i}h)^2-\sum_{k=1}^{k<\frac{i}{2}}\partial_{x}^{2k}h\partial_{x}^{2i-2k}h\right)+U(h)\right]_{n-\frac{3}{2}}^{n+\frac{1}{2}}\\
&+j_{n-\frac{1}{2}}(B_{n}+B^{\prime}_{n-1})+\mu_{n-\frac{1}{2}}[h]_{n-\frac{3}{2}}^{n+\frac{1}{2}}.
\end{split}
\end{equation}

Because of the defintion of $j$, $\dot{x}_{n}[h]_{n-\frac{1}{2}}^{n+\frac{1}{2}}=j_{n+\frac{1}{2}}-j_{n-\frac{1}{2}}$.
Finding $j_{n-\frac{1}{2}}$ from \eqref{eqint3} and $j_{n+\frac{1}{2}}$ from the same equation with
$n\to n+1$, we finally obtain the kink dynamics
\begin{equation}\label{kinkdynamicsC}
\begin{split}
\dot{x}_{n}[h]_{n-\frac{1}{2}}^{n+\frac{1}{2}}=&\frac{1}{B_{n+1}+B^{\prime}_{n}}\times\\
&\left(-\dot{x}_{n}A_{n}-\dot{x}_{n+1}A_{n+1}+\left[\sum_{i}a_{2i}\left(\frac{(-1)^{i-1}}{2}(\partial_{x}^{i}h)^2-\sum_{k=1}^{k<\frac{i}{2}}\partial_{x}^{2k}h\partial_{x}^{2i-2k}h\right)-U(h)\right]_{n-\frac{1}{2}}^{n+\frac{3}{2}}-\mu_{n+\frac{1}{2}}[h]_{n-\frac{1}{2}}^{n+\frac{3}{2}}\right)\\
&-\frac{1}{B_{n}+B^{\prime}_{n-1}}\times\\
&\left(-\dot{x}_{n-1}A_{n-1}-\dot{x}_{n}A_{n}+\left[\sum_{i}a_{2i}\left(\frac{(-1)^{i-1}}{2}(\partial_{x}^{i}h)^2-\sum_{k=1}^{k<\frac{i}{2}}\partial_{x}^{2k}h\partial_{x}^{2i-2k}h\right)-U(h)\right]_{n-\frac{3}{2}}^{n+\frac{1}{2}}-\mu_{n-\frac{1}{2}}[h]_{n-\frac{3}{2}}^{n+\frac{1}{2}}\right).
\end{split}
\end{equation}

We now estabish the relations
\begin{equation}
\begin{split}
[h]_{n-\frac{1}{2}}^{n+\frac{1}{2}} &= (-1)^{n}2h\ss{m} \\
B_{n+1}+B^{\prime}_{n}&=\int_{n+\frac{1}{2}}^{n+\frac{3}{2}}\mathrm{d}x~\partial_{x}h(x-x_{n+\frac{1}{2}})-\int_{n-\frac{1}{2}}^{n+\frac{1}{2}}\mathrm{d}x~\partial_{x}h(x_{n+\frac{1}{2}}-x)\\
&=\int_{n+\frac{1}{2}}^{n+\frac{3}{2}}\mathrm{d}x~h_{n+\frac{3}{2}}+\int_{n-\frac{1}{2}}^{n+\frac{1}{2}}\mathrm{d}x~h_{n-\frac{1}{2}}-\int_{n-\frac{1}{2}}^{n+\frac{3}{2}}\mathrm{d}x~h\\
&\simeq(-1)^{n+1}2h\ss{m} (x_{n+1}-x_{n}).\nonumber
\end{split}
\end{equation}
Then, reminding that $\mu_{n+\frac{1}{2}}=\sum_{i}a_{2i}\partial_{x}^{2i}h\mid_{n+\frac{1}{2}}-U^{\prime}(h_{n+\frac{1}{2}})$ is very small for a membrane near the equilibrium and $[h]_{n-\frac{1}{2}}^{n+\frac{3}{2}}$ is very small in the limit of distant kinks, 
Eq.~\eqref{kinkdynamicsC} becomes
\begin{equation}
\begin{split}
\dot{x}_{n}&=\frac{-1}{4h\ss{m}^2}\left\{\frac{1}{x_{n+1}-x_{n}}\left(-\dot{x}_{n}A_{n}-\dot{x}_{n+1}A_{n+1}+\left[\sum_{i}a_{2i}\left(\frac{(-1)^{i-1}}{2}(\partial_{x}^{i}h)^2-\sum_{k=1}^{k<\frac{i}{2}}\partial_{x}^{2k}h\partial_{x}^{2i-2k}h\right)-U(h)\right]_{n-\frac{1}{2}}^{n+\frac{3}{2}}\right)\right.\\
&\left.+\frac{1}{x_{n}-x_{n-1}}\left(-\dot{x}_{n-1}A_{n-1}-\dot{x}_{n}A_{n}+\left[\sum_{i}a_{2i}\left(\frac{(-1)^{i-1}}{2}(\partial_{x}^{i}h)^2-\sum_{k=1}^{k<\frac{i}{2}}\partial_{x}^{2k}h\partial_{x}^{2i-2k}h\right)-U(h)\right]_{n-\frac{3}{2}}^{n+\frac{1}{2}}\right)\right\}
\end{split}
\end{equation}
or, collecting $\dot{x}_{n}$ terms in the right-hand side,
\begin{equation}
\dot{x}_{n}=\frac{1}{4 h\ss{m}^{2}\l_{n}\l_{n-1}-A_{n}(\l_{n}+\l_{n-1})}
\left\{\l_{n-1}\left[\dot{x}_{n+1}A_{n+1}+f(\l_{n+1}, \l_{n-1})\right]
+\l_{n}\left[\dot{x}_{n-1}A_{n-1}+f(\l_{n}, \l_{n-2})\right]\right\}
\tag{\ref{eq_app}}
\end{equation}
where$A_n$ and $f(x,y)$ are defined in the main text respectively Eqs.~\eqref{eq_An} and \eqref{eq_fxy}.

\section{Numerical schemes}
\label{app_num}

We integrate the full dynamics of various PDE's
on a one-dimensional lattice with periodic boundary conditions. 
Space-derivatives are calculated using a finite-size difference scheme with
discretization $dx=0.2$. The time integration
is performed using an explicit Euler scheme.
The time-step depends on the equation
to be solved. The most stringent case is CH6, where 
must be $dt=10^{-6}$. Therefore, we have used 
this value of $dt$ for all simulations.

Initial conditions are built from stationary kink profiles, 
which are known analytically for TDGL and CH. 
For the fourth order case, TDGL4 and CH4, these kink profiles are
obtained numerically from steady-state solutions with isolated kinks. 
To build a kink-antikink pair at $x_{1}$ and $x_{2}$, 
we use the profile of the stationary kink $\hk(x-x_1)$ 
for $0\leq x\leq({x_{1}+x_{2}})/{2}$, 
and the profile of the stationary antikink, $-\hk(x-x_2)$, otherwise.

For the implementation of kink dynamics we also use an explicit Euler scheme
with $dt=10^{-2}$. The main difficulty comes from the annihilation of a kink-antikink pairs. 
Indeed, since kink models are designed to by quantitatively
accurate at long inter-kink distances only, 
they are not necessarily accurate, or even well defined
at short distances. However, kinks and antikinks
merge and annihilate rapidly in the full dynamics when their
separation is smaller than the size of the kink core.
We therefore generically use a cutoff inter-kink distance $a$ below which
kinks and anti-kinks spontaneously annihilate in kink dynamics,
using the following procedure.
If the separation between the $n$-th and $(n+1)$-th kinks 
is smaller than $a$ at time $t+dt$, 
we define the collision time-step $dt\ss{ann}=(x_{n+1}(t)-x_{n}(t))/(\dot x_{n}(t)-\dot x_{n+1}(t))$
from a simple linear extrapolation.
We then integrate the dynamics of all kinks up to $t+dt\ss{ann}$,
and erase the two kinks at $x_n$ and $x_{n+1}$.

In the non-conserved case, the full dynamics is not much affected
by the choice of the cutoff, and we chose $a=0$ in the simulations
presented in the main text. 
In the conserved case, the denominator $(4h\ss m)^{2}\ell_{n}\ell_{n-1}-A(\ell_{n}+\ell_{n-1})$ 
in Eq.(\ref{v_n-ch4}) vanishes for positive $\ell_{n}$ and $\ell_{n-1}$. 
Assuming that $\ell_n\sim\ell_{n-1}$ just before the collision,
we find that the critical value of the interkink distance to keep
the dynamics well defined is $\ell_c=2/(2h_m)^2$.
Using the quartic potential $U=-h\ss{m}^2h^2/2+h^4/4$,
with $h_m=0.9$, we find $A\approx1.86$, leading
to $\ell_c\approx 1.15$. We find that
our numerical scheme is stable for $a\geq 1.3$,
which is consistent with the expected constraint $a>\ell_c$.
In order to ensure strong stability, 
we have performed most simulations with a slightly larger
value of $a=1.5$. 

\bibliography{kinks}

\end{document}